\documentclass[notitlepage,twocolumn,prl,tightenlines,nofootinbib,superscriptaddress,showpacs]{revtex4}

\usepackage{amsmath}
\usepackage{amssymb,amsfonts}
\usepackage{bm}
\usepackage[mathcal]{euscript}
\usepackage{graphicx}
\usepackage{epsfig}
\usepackage{color}

\begin{document}

\title{The building blocks of dynamical heterogeneities in dense granular media}

\author{R. Candelier}
\affiliation{SPEC, CEA-Saclay, 91 191 Gif-sur-Yvette, France}
\author{O. Dauchot}
\affiliation{SPEC, CEA-Saclay, 91 191 Gif-sur-Yvette, France}
\author{G. Biroli}
\affiliation{Institut de Physique Th{\'e}orique, CEA, IPhT, F-91191 Gif-sur-Yvette, France and CNRS, URA 2306}

\begin{abstract}
We investigate experimentally the connection between short time dynamics and long time dynamical heterogeneities within a dense granular media under cyclic shear. We show that dynamical heterogeneities result from a two timescales process. Short time {\it but already collective} events consisting in clustered cage jumps concentrate most of the non affine displacements. On larger timescales such clusters appear aggregated both temporally and spatially in avalanches which eventually build the large scales dynamical heterogeneities. Our results indicate that facilitation plays an important role in the relaxation process although it does not appear to be conserved as proposed in many models studied in the literature.   
\end{abstract}

\maketitle

Mechanically driven grains exhibit a dramatic slowing down of their dynamics when their volume fraction is increased above a certain value. This phenomenon, generically called jamming transition, shares a lot of experimental features with the glass transition and, indeed, it has been suggested that they are both governed by similar underlying physical mechanisms \cite{Liu98}. Whether such mechanisms originate from an ideal transition of any kind remains however a matter of debate \cite{biroli2007np,mari2008,Berthier08}. One of the major recent advances in these fields has been the discovery of dynamic heterogeneity (DH). Experimental and numerical works have shown that the dynamics become increasingly correlated in space approaching the glass and the jamming transitions (see \cite{biroli2007asm} for a recent review, and \cite{marty2005sac,dauchot2005dhc,abate2006aja,lechenaultEPLI,lechenaultEPLII} more specifically for granular media). This clearly shows that the slowing down of the dynamics is related to a collective phenomenon, possibly to a true phase transition. Different theories have been developed in order to explain quantitatively this phenomenon. The crucial last missing piece consists in understanding what is the underlying mechanism leading to dynamic heterogeneity and, hence, responsible for the slow relaxation. Many different possible origins have been highlighted in the literature: dynamic facilitation \cite{garrahan2002prl}, soft modes \cite{brito2007hdm,harrowell2008irs}, proximity to a mode coupling transition \cite{donati2002tnl,biroli2004dls}, growing amorphous order \cite{lubchenko2006tsg, biroli2008tsg}, etc. At this stage, it is therefore crucial to perform detailed studies aimed at unveiling what are the building blocks of DH.

The aim of this letter is to perform such type of analysis for a granular system close to its jamming transition. Our starting point consists in identifying the elementary irreversible relaxation processes, that we shall call cage jumps in reference to the well known interpretation of the slowing down of the dynamics in term of caging \cite{biroli2007asm}. Our analysis shows that DH is the result of two processes taking place on different timescales. On short timescales, clustered cage jumps concentrate most of the non affine displacements. On larger timescales such clusters, that are already collective events, aggregate both temporally and spatially in avalanches and ultimately build the large scales dynamical heterogeneities. We find that dynamic facilitation \cite{fredrickson1984,garrahan2002prl} clearly plays a major role in the development of the avalanche process although it seems to be irrelevant in triggering it. A detailed discussion of our findings on the basis of the current theoretical literature is presented in the conclusion.


The experimental setup, the same as in \cite{marty2005sac,dauchot2005dhc} consists in an horizontal monolayer of about 8300 bi-disperse steel cylinders of diameter 5 and 6 mm in equal proportions quasi-statically sheared at constant volume fraction $\phi=0.84$. The shear is periodic, with an amplitude $\theta_{max}=\pm5^{\circ}$. A high resolution camera takes images each time the system is back to its initial position $\theta=0^{\circ}$. Both the camera resolution and a better control of the lightening uniformity now allow the tracking of $N=4055$ grains in the center of the device, without any loss. A typical experiment lasts 10 000 cycles. We choose the time unit to be one back and forth cycle, and the length unit to be the diameter of small particles. Redoing the same analysis as in previous studies~\cite{marty2005sac,dauchot2005dhc} we observe that: (i) the dynamics is isotropic, subdiffusive at short times and diffusive at long times; subdiffusion stems from the trapping of the particles within cages of size $\sigma_c=0.1$, a value slightly smaller than in~\cite{marty2005sac,dauchot2005dhc} presumably because of small changes in packing fraction and/or shear amplitude; (ii) introducing 
\begin{equation}
Q_{p,t}(a,\tau) = \exp \left( - \frac{||\Delta \vec{r}_p(t,t+\tau)||^2}{2a^2} \right),
\label{def_de_Q}
\end{equation}
\noindent
where $\Delta \vec{r}_p(t,t+\tau)$ is the displacement of the particle $p$ between $t$ and $t+\tau$ and $a$ is a probing length scale. The computation of the four points correlation function $\chi_4(a,\tau)=N (<Q_t(a,\tau)^2>-<Q_t(a,\tau)>^2)$, where $Q_t(a,\tau)=\frac{1}{N} \sum_p Q_{p,t}(a,\tau)$ reveals that the dynamical correlation length is maximal for $\tau^*=720$ and $a^*=0.15$.


In the present study, we first segment the trajectories in separated cages introducing a novel algorithm. Consider a trajectory $S(t)_{t \in [0,T]}$ on a total time $T$ and split it at an arbitrary cut time $t_c$ into two sets of successive points : $S_1$ for $t_1 \in [0,t_c]$ and $S_2$ for $t_2 \in ]t_c,T]$. Then we measure how well separated are the two sets of points:
\begin{equation}
p(t_c) = \xi(t_c).\left(<d_1(t_2)^2>_{t_2 \in S_2}.<d_2(t_1)^2>_{t_1 \in S_1}\right)^{1/2}
\label{def_de_p}
\end{equation}

\begin{figure}[t] 
\centering
\includegraphics[width=0.49\columnwidth]{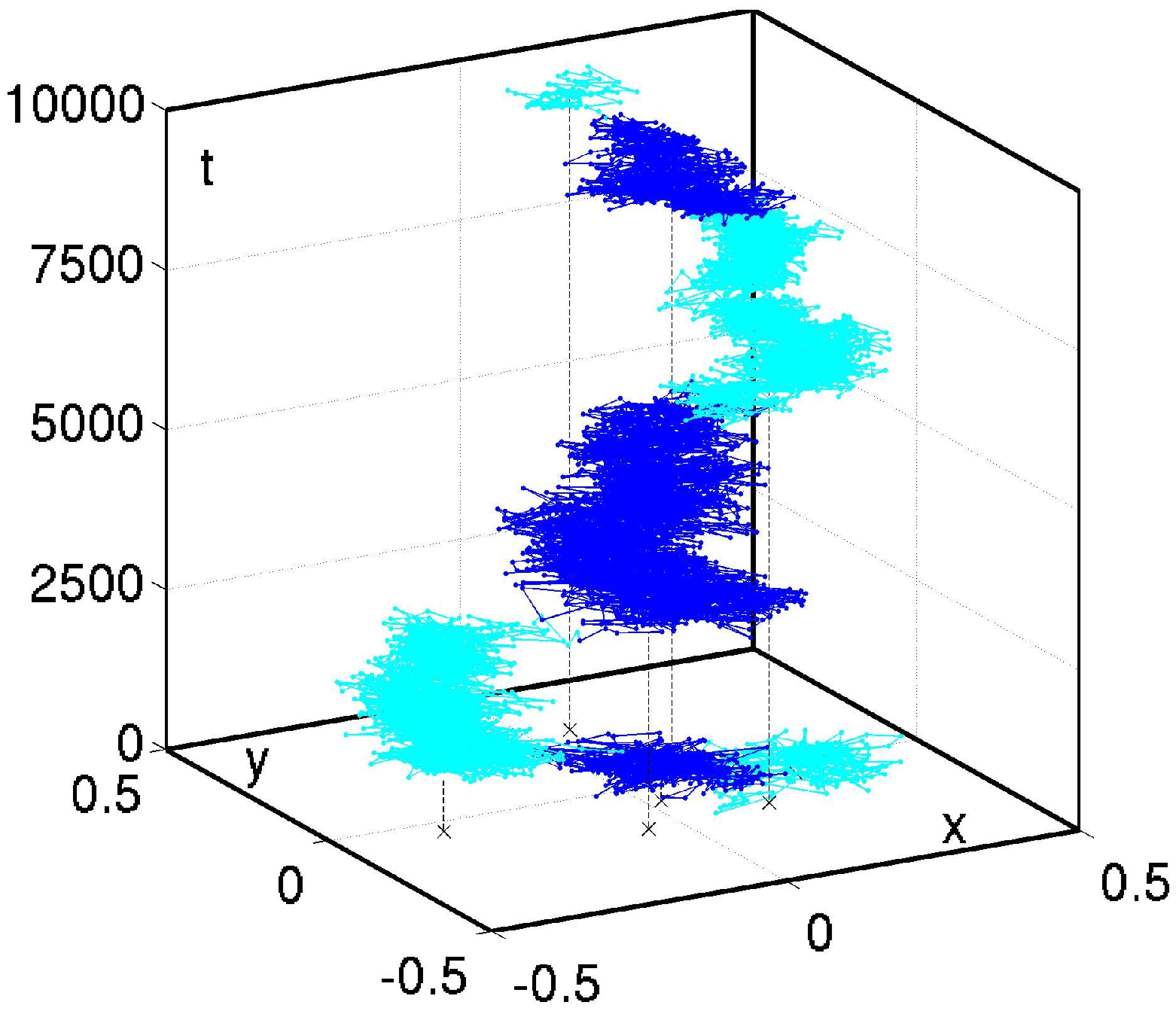}
\includegraphics[width=0.49\columnwidth]{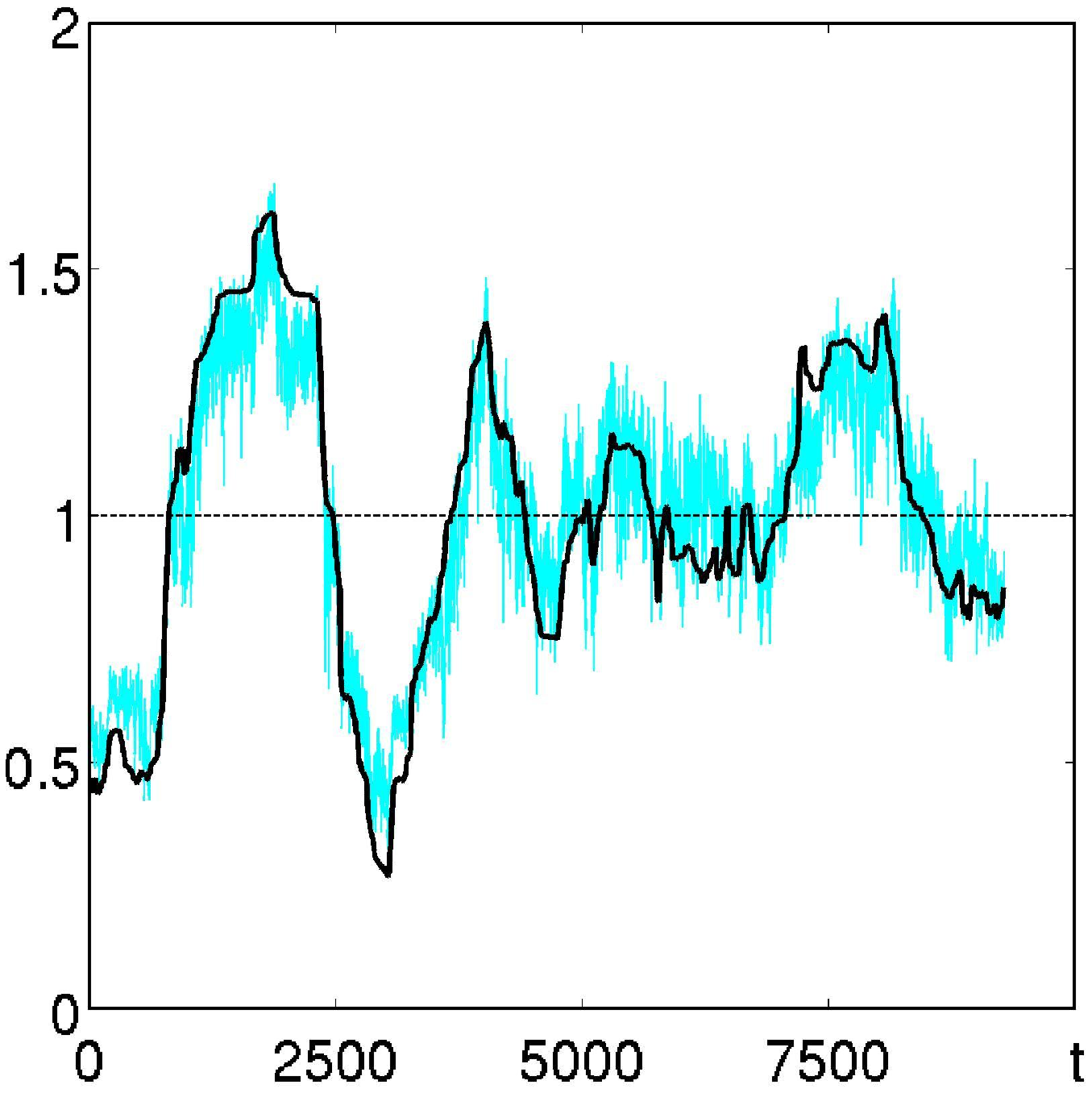}
\caption{\textbf{Left:} 3D visualization of the trajectory of a single particle. The color changes every time the algorithm detects a cage jump. \textbf{Right:} Comparison between the relative averaged relaxation $Q_t(a^*,\tau^*)/<Q_t>_t$ (in gray/cyan online) and the relative percentage $P_t(\tau^*)/<P_t>_t$ of particles that haven't jumped between $t$ and $t+\tau^*$ (in dark), $\tau^*=720$.}
\label{fig:cages}
\vspace{-0.25cm}
\end{figure}
\begin{figure*}[t!] 
\centering
\includegraphics[width=0.68\columnwidth]{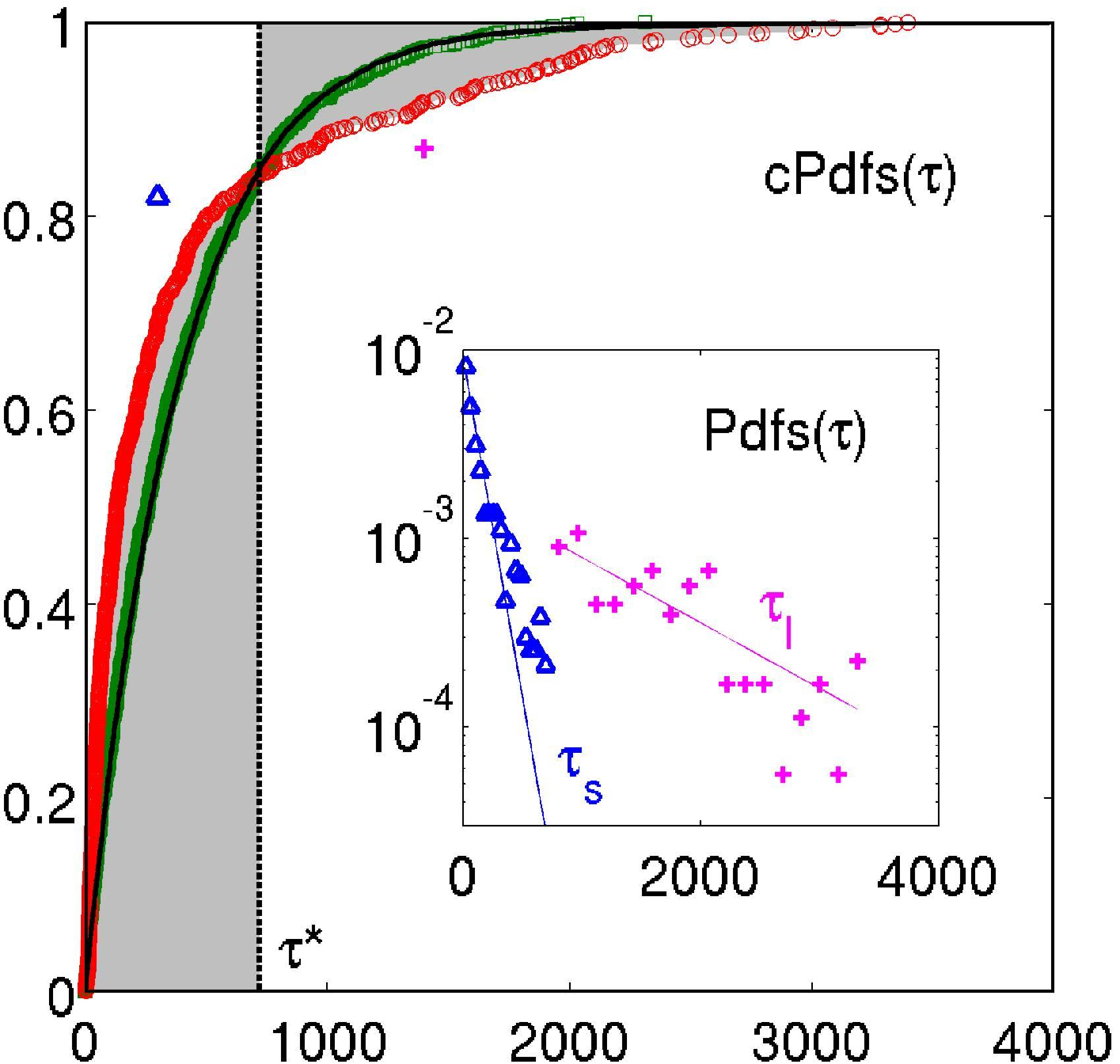}
\includegraphics[width=0.65\columnwidth]{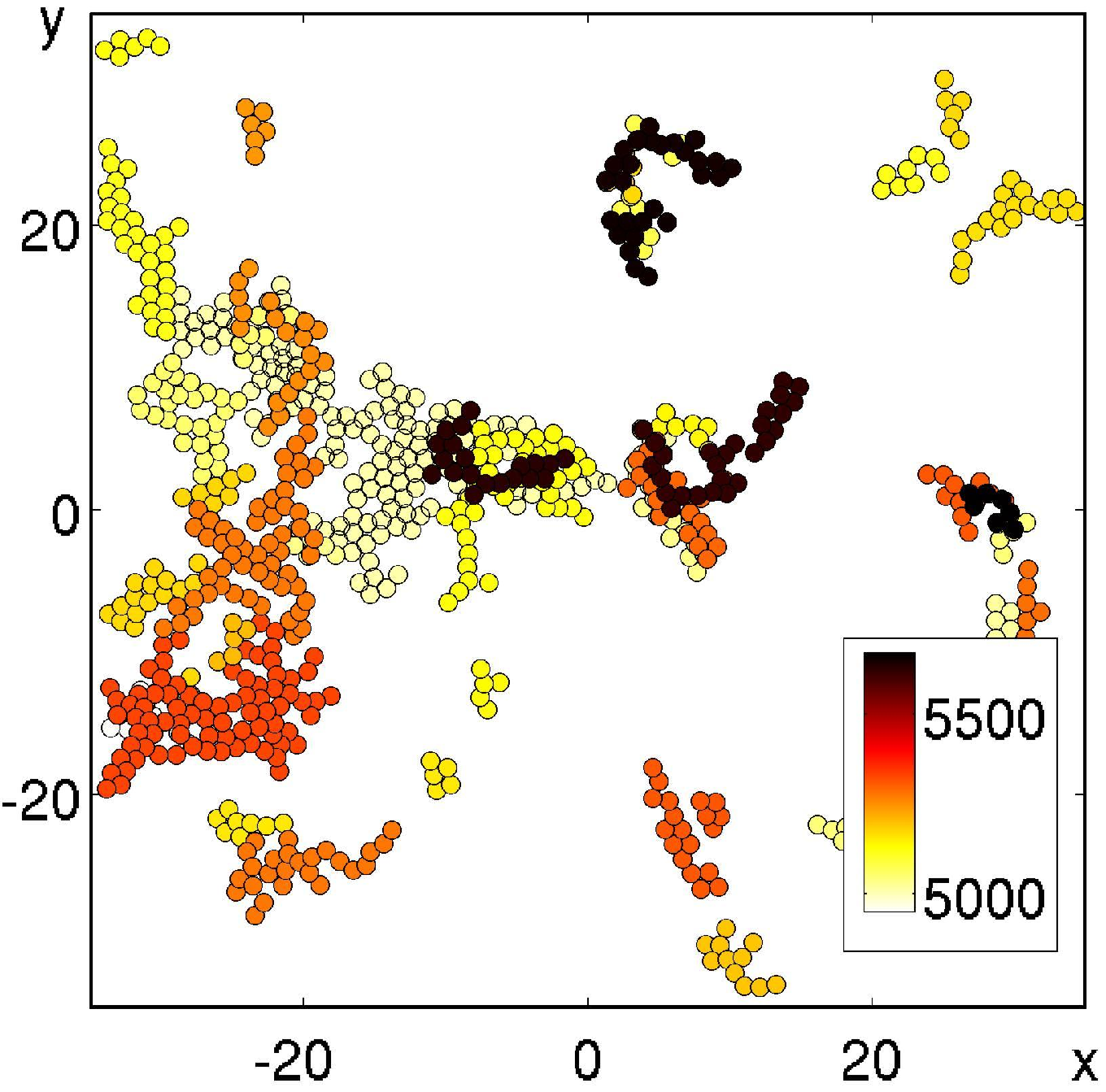}
\includegraphics[width=0.65\columnwidth]{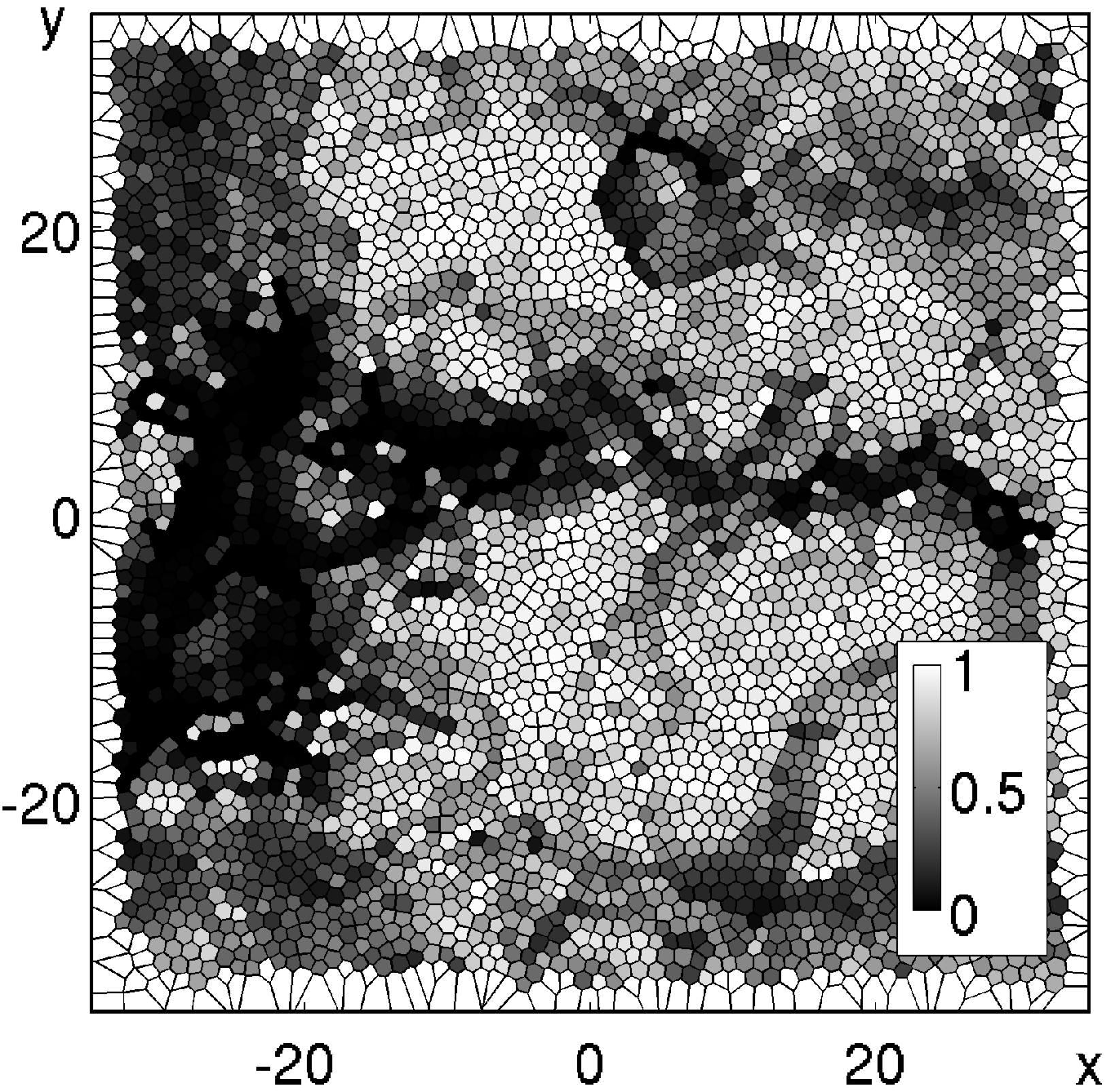}
\caption{\textbf{Left:} Cumulated probability distribution of the duration between spatially adjacent clusters (color online). Experimental data are in red circles while green squares stand for a set of equal cardinality generated from an exponential distribution with the same mean value. The black curve is the analytical version of this same distribution. The actual distribution and the exponential ones cross at the lag time $\tau^*=720$. Inset: Distribution of the lags below $\tau^*$ (blue triangles) and above (magenta crosses). \textbf{Middle:} Spatial location of successive clusters of cage jumps. Colors correspond to the time at which clusters occur. \textbf{Right:} Spatial field of the two point correlation function $Q_{p,t}(a^*,\tau^*)$.}
\label{fig:avalanches}
\vspace{-0.5cm}
\end{figure*}

\noindent where $d_k(t_i)$ is the distance between the point at time $t_i$ and the center of mass of the subset $S_k$. The average $<>_{S_k}$ is computed over the subset $S_k$. $\xi(t_c)=\sqrt{t_c/T*(1-t_c/T)}$, the standard deviation of the number of steps in a given set for a uniformly distributed process is the natural normalisation that eliminates the large fluctuations arising when $t_c$ is too close to the bounds of $[0,T]$. We define a cage jump at $t_c$ when $p(t_c)$ is maximal. The procedure is then repeated iteratively for every sub-trajectory until $p_{max}(t_c)<\sigma_c^2$. Fig.~\ref{fig:cages}-left illustrates how, using the above algorithm, we successfully segment the trajectories into cages separated by jumps. Cage jumps are defined within a resolution of 10 cycles. A direct and important observation is that the distribution of the time spent in each cage is exponential and characterized by an average "cage time" $\tau_c=1160$. For comparison, $Q_t(a^*,\tau=1000)\simeq 0.5$ \cite{dauchot2005dhc}. This means that in average a particle jumps only once on the timescale over which $Q_t(a^*,\tau)$ relaxes. Fig.~\ref{fig:cages}-right displays the relative values of $Q_t(a^*,\tau^*)$ together with $P_t(\tau^*)$ the relative percentage of particles that have not jumped during $\tau^*$. The correlation is straightforward~: the bursts of cage jumps caught by the algorithm are responsible for the major relaxation events of the system. Anticipating on the following, one can also check that the cage jumps detected by the algorithm are also exactly located in the areas where the decorrelation is maximal (compare fig.~\ref{fig:avalanches} middle and right). 

\begin{figure}[t] 
\center
\includegraphics[width=0.49\columnwidth]{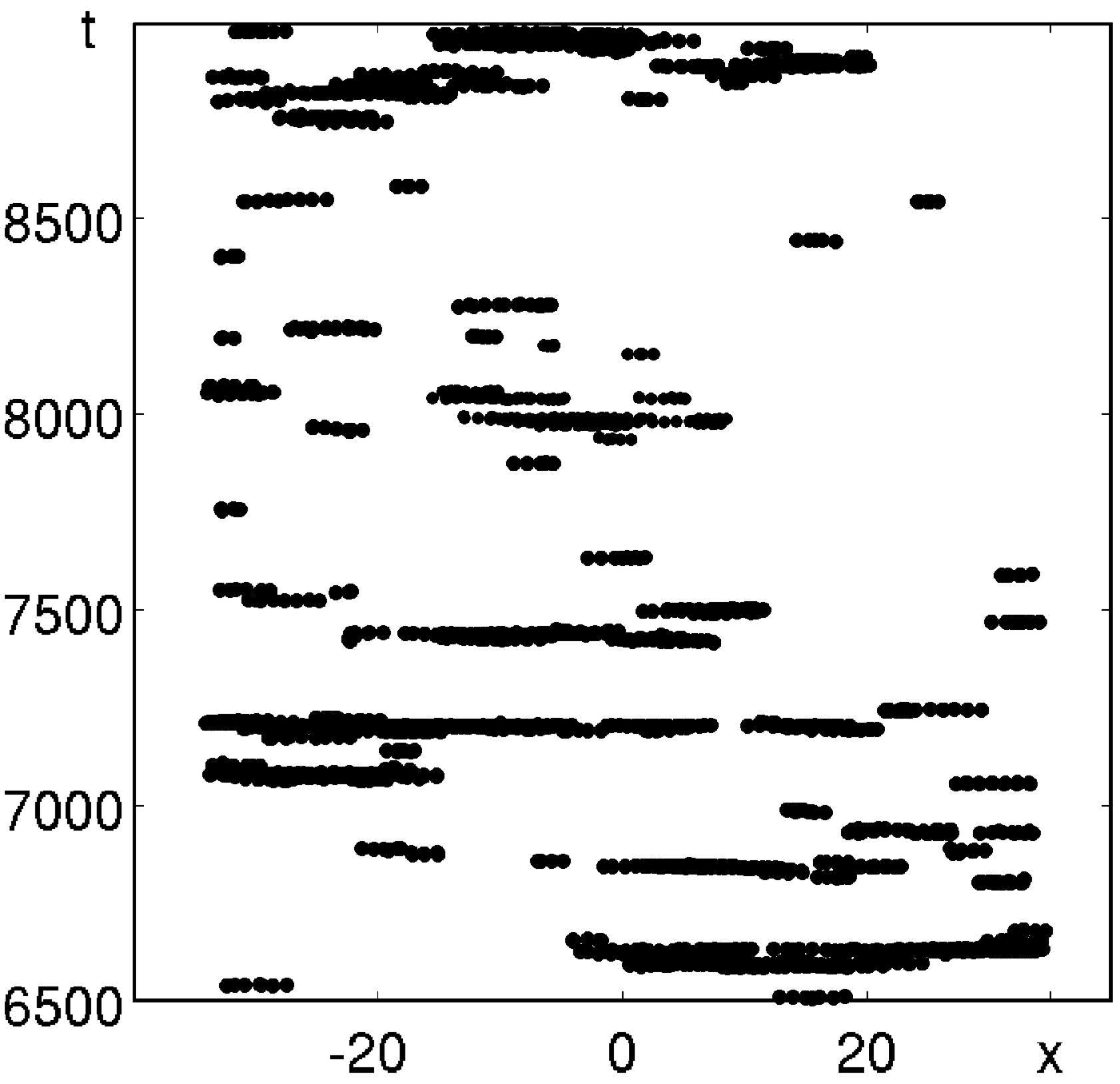}
\includegraphics[width=0.49\columnwidth]{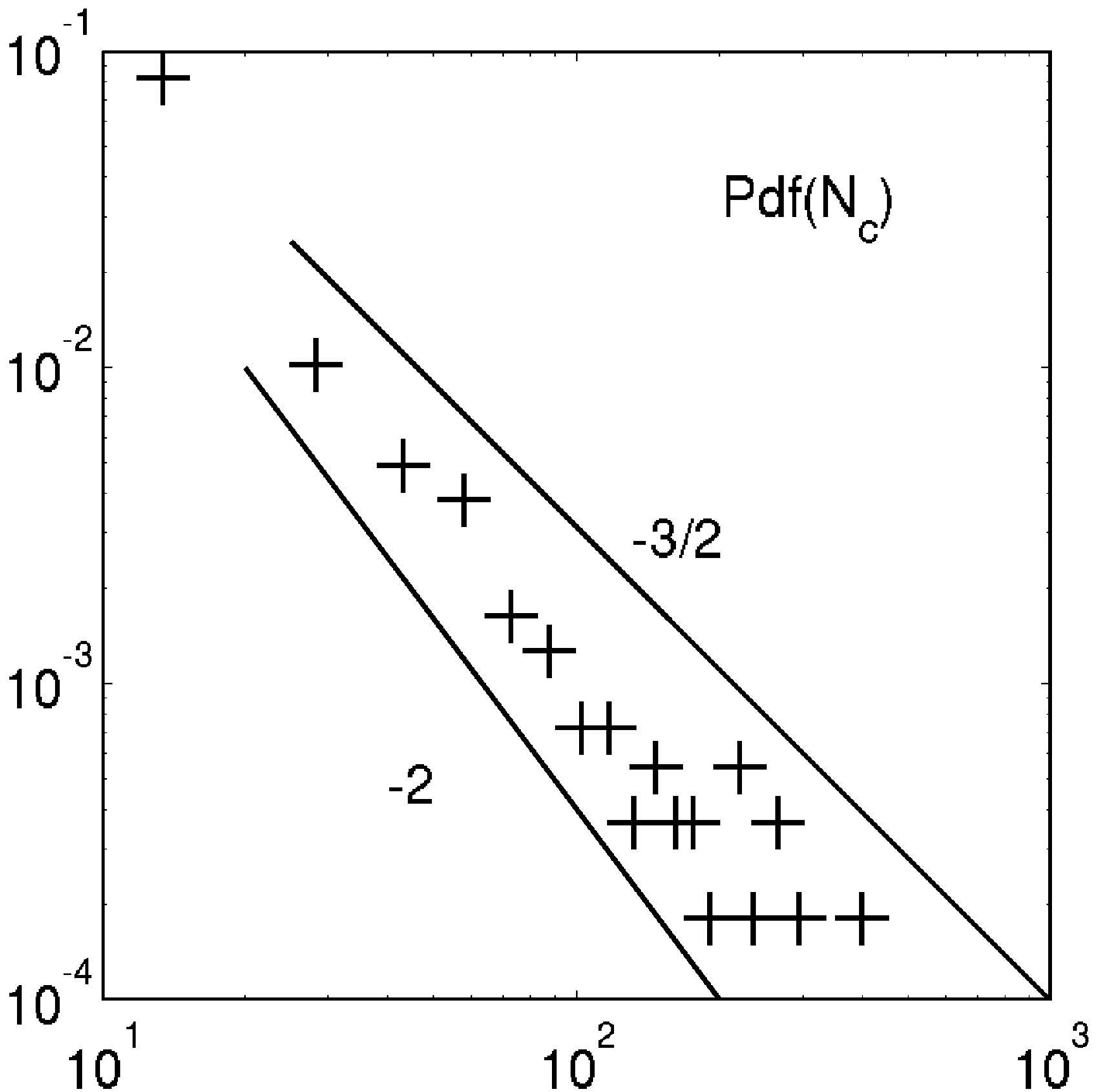}
\caption{\textbf{Left:} Spatio-temporal position of the cage jumps; only one direction in space is shown (x-axis). Each point represents a cage jump. The very flat clouds of points are clusters of \textit{collective} and \textit{instantaneous} cage jumps. \textbf{Right:} Probability distribution of clusters sizes.}
\label{fig:clusters}
\vspace{-0.5cm}
\end{figure}


Fig.~\ref{fig:clusters}-left reveals that cage jumps occur intermittently both in space and time. There are very long intervals without a jump in a whole region of space separated by sudden and {\it collective} relaxation events. When clustering the cage jumps which are adjacent in space -- neighbouring particles -- and time -- separated by less than the jump resolution, 10 cycles --  one can extract two important features. The duration of these clusters follows an exponential distribution with an average value which remains small, typically of the order of 10 cycles. On the contrary, cage jumps are not isolated in space: the cluster size distribution has very fat tails. In the regime of sizes experimentally available, it is well described by a power law $\rho(N_c)\simeq N_c^{-\alpha}$, where $N_c$ is the number of grains within a cluster and $\alpha \in [3/2,2]$ (see fig.~\ref{fig:clusters}-right). Experimentally the average cluster size equals $18$ and has a standard deviation of $34$. We now compute the square difference between the actual local deformation around a grain $i$, and the one it would have if it were in a region of uniform strain $\underline{\underline{{\varepsilon}}}$: 
\begin{equation}
D_i^2(t,\tau) = \sum_{j}\left( \vec{r}_{ij}(t+\tau)-\underline{\underline{{\varepsilon}}}.\vec{r}_{ij}(t)\right)^2,
\end{equation}
%
\begin{figure}[b!] 
\center
\includegraphics[width=0.50\columnwidth]{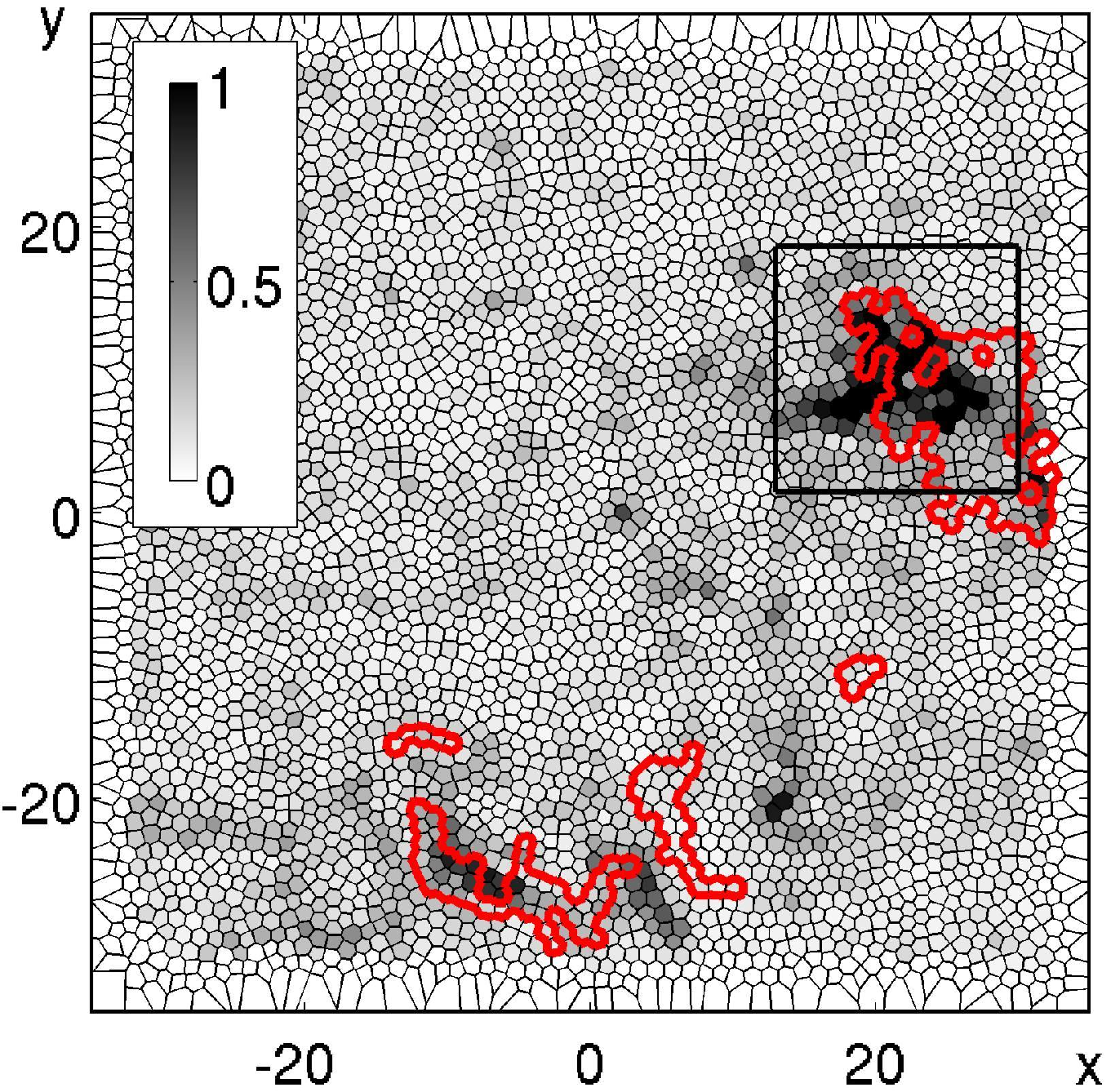}
\includegraphics[width=0.485\columnwidth]{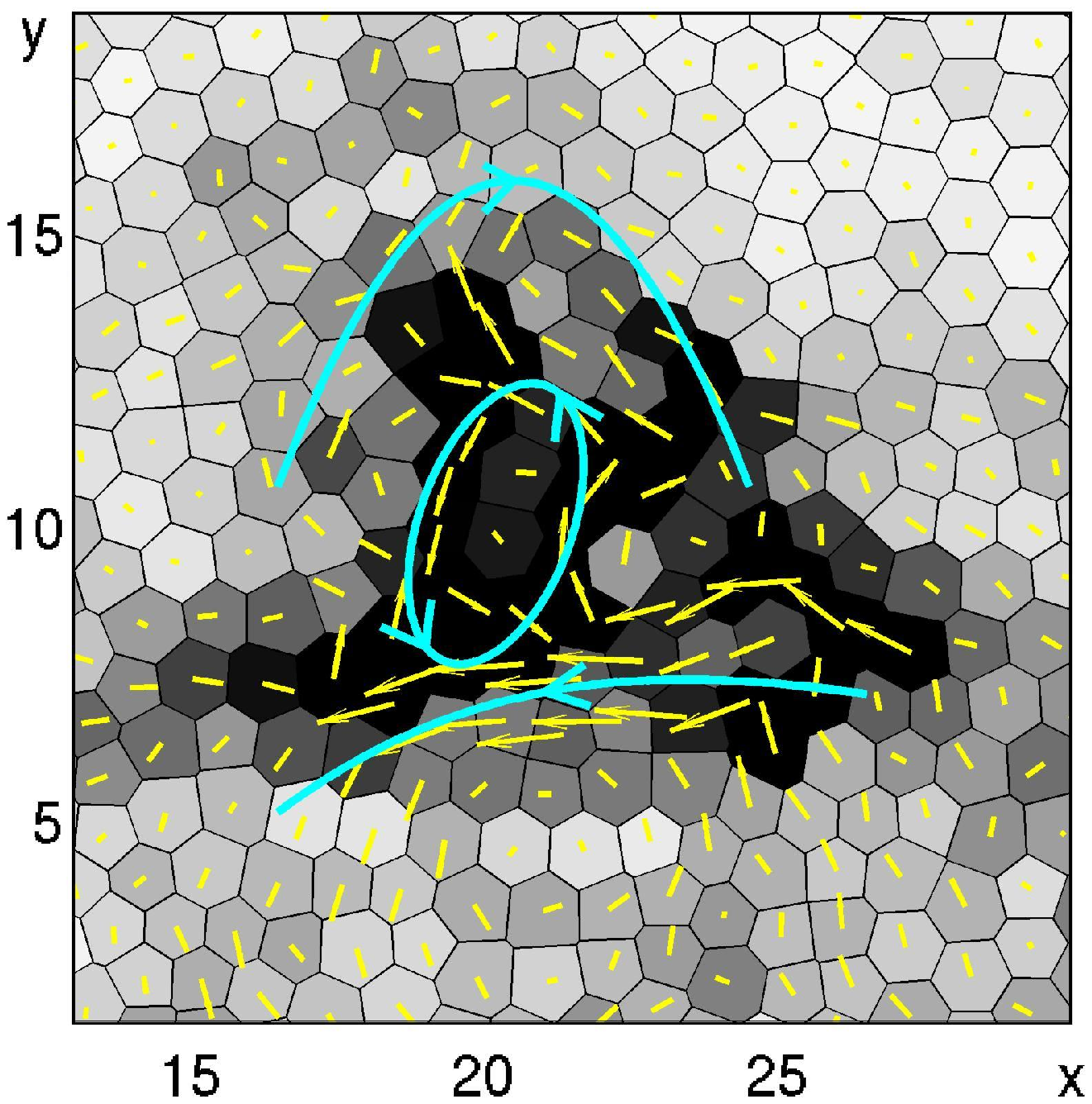}
\caption{\textbf{Left:} Clusters of cage jumps concentrate the highly non-affine domains : the distance to affinity parameter $\Delta(t,\tau)$ (in levels of gray) is compared to the location of clusters of collective cage jumps (with red boundaries) ($\tau$=30). \textbf{Right:} Zoom on a highly non-affine region (box on the left figure). The displacements of the particles, magnified by a factor 2, are in light grey (yellow online). For convenience, the main streams creating intense local shears are eye-guided.}
\label{fig:stz}
\vspace{-0.0cm}
\end{figure}

\noindent where the index $j$ runs over the neighbouring grains of reference grain $i$ and $\vec{r}_{ij}(t)=\vec{r}_{j}(t)-\vec{r}_{i}(t)$. $\Delta_i^2(t,\tau)=\mathrm{ Min}_{\underline{\underline{{\varepsilon}}}}\left(D_i^2(t,\tau)\right)$ is the local deviation from affine deformation during the time interval $\tau$ (see \cite{falk1998dov} for details). We observe (fig.\ref{fig:stz}) that the clusters of cage jumps concentrate the highly-non affine deformations and can be identified as the elementary irreversible events of the dynamics.


We shall now unveil how the above short term events build up large collective relaxation on long time scales. The heavy tails in the distribution of the cluster sizes (fig.~\ref{fig:clusters}-right) suggests that the collective cage jumps aggregate into some kind of avalanche process : a first cluster triggers the apparition of successive bursts nearby shortly after, which in turn trigger other nearby bursts. Such avalanches would provide a natural mechanism for the formation of the long term dynamical heterogeneities, as we shall see now. Fig.~\ref{fig:avalanches}-left compares the cumulative distribution ($cPdf$)of the lag times between adjacent clusters (red circles) to that of independent events following a Poissonian process with the same average lag time (dark line). Both cumulative distributions intersect at a lag time corresponding precisely to the timescale of the dynamical heterogeneities $\tau^*$: compared to the Poissonian process, there is an excess of short lag times when $\tau<\tau^*$ 
i.e. $\mathrm{Prob}(lag<\tau)=cPdf(\tau)$ is larger than for the Poissonian process
, and an excess of large lag times when $\tau>\tau^*$ 
i.e.  $\mathrm{Prob}(lag>\tau)=1-cPdf(\tau)$ is again larger than for the Poisonian process 
, leading to an under representation of intermediate lag times\footnote{Note that the Poissonian distibution computed from a randomly generated data set with the same cardinality and the same average (green squares) is identical to the analytical curve, excluding any finite size effect in the above observation.}. In the inset, one can see the two lag time distributions corresponding to events separated by respectively less and more than $\tau^*$; these exponential distributions reveals two very different typical decay times ($\tau_s=120$ and $\tau_l=1190$). This separation of times underlines the aggregation of the clusters of cage jumps into separated avalanches. The short time scale $\tau_s$ corresponds to the delay between two successive events within a given avalanche, whereas the long one $\tau_l$ is the time separating two avalanches at a similar location. $\tau_l$ nicely corresponds to the typical cage time of individual particles $\tau_c=1160$, indicating that almost no particle jumps twice within the same avalanche. Spatially, note that the minimal distance between avalanches\footnote{We compute the distance between avalanches as the minimal distance between all the couples of clusters separated of a lag time less than 2$\tau_s$ and belonging to different avalanches.} points toward an average distance of $27$ and a standard deviation of $14$, indicating a clear spatial separation between avalanches. Also, the fractal dimension of clusters $d_F$ gives a geometrical characterization of the structure of the dynamically correlated regions. Within the statistical accuracy, $d_F$ increases from 1.3 towards 2 during the aggregation process. Thus, as in numerical studies on glass-forming liquids \cite{donati1999scm,appignanesi2006prl}, we find that dynamically correlated regions becomes thicker on larger timescales. Finally, selecting a time interval of length $\tau^*$, initiated at the beginning of a given avalanche, fig.~\ref{fig:avalanches} compares the spatial organization of the clusters in the avalanche and the local relaxation of the system as measured by the field $Q_{p,t}(a^*,\tau^*)$. The correspondence is very good: the aggregation of all the clusters within an avalanche is ultimately building a large decorrelation area, also seen on the correlation function $Q_{p,t}(a^*,\tau^*)$. More interestingly, each cluster in fig.~\ref{fig:avalanches}-middle is colored according to a color gradient corresponding to the time at which it occurs, thereby underlining the way a first cluster of cage jumps has given rise to successive neighbouring clusters.


To summarize, we have identified a two timescales process that give rise to dynamical heterogeneities (DH) and is responsible for macroscopic relaxation. At short times, the particles collectively jumps within clusters whose sizes are very largely distributed. These clustered jumps trigger other ones nearby within an avalanche process.  The lifetime of such avalanches is much smaller than the timescale between two avalanches in a similar location or, analogously, between succesive cage jumps of a given grain. DH are strongest on a timescale which corresponds to the crossover between these last two. It is interesting to discuss our results within the perspectives provided by current theoretical approaches. Dynamic facilitation (DF) is one mechanism put forward to explain slow and glassy dynamics. Theoretical approaches based on DF usually focus on Kinetically Constrained Models (KCM) \cite{ritort2003gdk,garrahan2002prl,toninelli2006jpa}. They are characterized by a common mechanism leading to slow dynamics: relaxation is due to mobile facilitating regions that are rare and move slowly across the system.  Here, we find a dynamics characterized by avalanches inside which clusters are facilitating each other. It is important to remark that the fraction of particles relaxing because of facilitation, i.e. belonging to a cluster but the first one (in time) of an avalanche, is close to $0.85$. However, in our system facilitation is not conserved as in KCMs since the first cluster of an avalanche is far from any other possible facilitating region. Why then particles jump in the first cluster of an avalanche? This is hardly a pure random event since it is already a collective phenomenon clustered in space and time. Promising candidates to explain it are the so called soft modes or soft regions. It has been shown that for hard spheres close to jamming \cite{brito2007hdm} and for moderately supercooled liquids \cite{harrowell2008irs} a significant fraction of the dynamical evolution takes place along the soft modes and dynamic heterogeneity is strongly correlated with the softest regions. One can then conjecture that the first clusters of avalanches correlate with the softest regions of the system. The resulting scenario is a mixture of the one based on soft modes and the one based on DF: dynamical evolution starts from the softest regions but then propagate on larger lengthscales by dynamic facilitation. Note that the relationship between these two pictures has been also discussed recently in an analysis of a kinetically constrained model~\cite{garrahan2008preprint}. Still, without having computed the soft modes in a frictional packing one cannot eliminate other possible (maybe complementary) mechanisms such as hopping between local minima in energy landscape~\cite{doliwa2003pre}.
It is also interesting to remark that the Mode Coupling Theory of the glass transition is based on the emergence of soft modes and predicts~\cite{biroli2006prl}, as we find, that dynamical correlated structures thicken in time. Obviously, all such conjectures call for further investigation. From the experimental point of view, one would like to identify the soft modes and check their correlation with the clusters we identified. Repeating the present study in simulations of glass-forming liquids would be certainly very instructive. One could check whether the building blocks of DH are the same ones we identified for granular media. Finally, it would be interesting to know how the processes we identified evolved with density and in particular which of DF and soft modes becomes more important when increasing the density.

We would like to thank J.~P. Bouchaud, P. Harrowell, S. Aumaitre, F. Lechenault for helpful discussions as well as V. Padilla and C. Gasquet for technical assitance on the experiment.

\bibliography{biblio_glass}


\end{document}